\pdfoutput=1

\documentclass[sigconf]{acmart}

\usepackage{balance}

\usepackage{hyperref}

\hypersetup{
    colorlinks=true,
    linkcolor=blue,
    filecolor=magenta,      
    urlcolor=cyan,
    pdftitle={Overleaf Example},
    pdfpagemode=FullScreen,
    }

\AtBeginDocument{%
  \providecommand\BibTeX{{%
    \normalfont B\kern-0.5em{\scshape i\kern-0.25em b}\kern-0.8em\TeX}}}

\copyrightyear{2022} 
\acmYear{2022} 
\setcopyright{acmlicensed}\acmConference[IVA '22]{ACM International Conference on Intelligent Virtual Agents}{September 6--9, 2022}{Faro, Portugal}
\acmBooktitle{ACM International Conference on Intelligent Virtual Agents (IVA '22), September 6--9, 2022, Faro, Portugal}
\acmPrice{15.00}
\acmDOI{10.1145/3514197.3549697}
\acmISBN{978-1-4503-9248-8/22/09}

\begin{document}

\title{Evaluating Data-Driven Co-Speech Gestures of Embodied\\ Conversational Agents through Real-Time Interaction}


\author{Yuan he}
\email{yuan-he@live.com}
\orcid{0000-0002-6126-6908}
\affiliation{%
  \institution{KTH Royal Institute of Technology}
  \streetaddress{Brinellvägen 8, 114 28}
  \city{Stockholm}
  \country{Sweden}
  \postcode{11428}
}

\author{Andr\'e Pereira }
\affiliation{%
  \email{atap@kth.se}
  \institution{KTH Royal Institute of Technology}
  \city{Stockholm}
  \country{Sweden}
}

\author{Taras Kucherenko}
\authornote{Work performed while at KTH}
\affiliation{%
  \email{tkucherenko@ea.com}
  \institution{SEED - Electronic Arts (EA)}
  \city{Stockholm}
  \country{Sweden}}


\begin{abstract}
Embodied Conversational Agents (ECAs) that make use of co-speech gestures can enhance human-machine interactions in many ways. In recent years, data-driven gesture generation approaches for ECAs have attracted considerable research attention, and related methods have continuously improved. Real-time interaction is typically used when researchers evaluate ECA systems that generate rule-based gestures. However, when evaluating the performance of ECAs based on data-driven methods, participants are often required only to watch pre-recorded videos, which cannot provide adequate information about what a person perceives during the interaction. To address this limitation, we explored use of real-time interaction to assess data-driven gesturing ECAs. We provided a testbed framework, and investigated whether gestures could affect human perception of ECAs in the dimensions of \textit{human-likeness}, \textit{animacy}, \textit{perceived intelligence}, and \textit{focused attention}. Our user study required participants to interact with two ECAs – one with and one without hand gestures. We collected subjective data from the participants’ self-report questionnaires and objective data from a gaze tracker. To our knowledge, the current study represents the first attempt to evaluate data-driven gesturing ECAs through real-time interaction and the first experiment using gaze-tracking to examine the effect of ECAs' gestures. 
\end{abstract}




\maketitle

\section{Introduction}
During interpersonal communication, we convey information not only through speech but also through non-verbal behaviors. It has been proven that non-verbal behaviors have critical impacts on interactions, such as revealing personality, emotions, and intimacy levels \cite{Knapp_Hall_Horgan_2012}. Therefore, numerous studies have focused on enabling virtual agents to mimic gestural human-to-human communicative behaviors in addition to verbal communication capabilities, thus enhancing the user's perception during interactions \cite{Feine_Gnewuch_Morana_Maedche_2019,Vinciarelli2012,M_expressiveagents}. Hand gestures, as one of the most commonly observed nonverbal behaviors, have received considerable research attention when designing ECAs \cite{Ravenet_Pelachaud_Clavel_Marsella_2018,Cassell_Vilhja}.

There have been two main approaches to enabling Embodied Conversational Agents (ECAs) to generate co-speech hand gestures. One approach is to identify the patterns produced by human cognition and behavior generation, summarize the correspondence between them, then model them accordingly, which we commonly refer to as \textit{rule-based methods} \cite{Cassell_Pelachaud_1994, Cassell2001, Huang_Mutlu_2013, Ravenet_Pelachaud_Clavel_Marsella_2018}. The other approach is to employ machine learning algorithms to build models from large amounts of human communication data, which we regard as \textit{data-driven methods} \cite{Levine2010, Yoon_Ko_Jang_Lee_Kim_Lee_2019, Alexanderson_Henter_Kucherenko_Beskow_2020, kucherenko2021moving}. Both types of methods have their pros and cons. Rule-based methods are more interpretable than data-based approaches, and the generated gestures can accurately reflect the agent's cultural background, gender, and other identifying characteristics. However, rule formulation requires theoretical research and a lot of human labor \cite{Chiu_Marsella_2011}. On the other hand, while data-driven approaches address the limitation of a lack of domain knowledge, assembling training data and mapping generated gestures to semantics remain the principal challenges.

Even though the techniques for generating gestures have been steadily optimized and improved, the methodology for evaluating their quality remains a research challenge. There is still no standardized method for quantitatively assessing gesture generation results \cite{Wolfert_Robinson_Belpaeme_2021}. Researchers have used objective measurements to evaluate a generative model's performance, such as comparing the joint velocity and position of generated and ground truth gestures \cite{Marmpena_Garcia_Lim_2020}. Although the required data can be obtained readily, objective evaluation cannot represent human interpretation of gesture performance. Therefore, user studies should also be conducted to evaluate generated gestures subjectively.

Using subjective evaluation methods to evaluate the effectiveness of gesture generation serves two principal purposes: First, to determine if the gestures generated by the proposed model can affect users' perceptions of interactions in some specific dimensions; and second, to compare the performances of different generative models regarding their perceived qualities in user experience. This study focused on serving the first purpose to evaluate the impact of one selected model on the user experience through real-time user interaction, but the experiment workflow developed in this research can also be applied to compare different models in the future.

Moreover, some objective indicators can be used to help compensate for the qualitative nature of subjective observations. For example, some studies have utilized eye-gazing behavior \cite{Chong_2018} and body posture \cite{Sanghvi_Castellano_Leite_Pereira_McOwan_Paiva_2011} to assess user engagement. We also implemented an eye tracker to observe the human's \textit{focused attention} when interacting with the ECA.

In most cases, rule-based systems are designed for existing platforms that feature interactive virtual agents (e.g. Greta \cite{pelachaud2015greta}). Therefore, evaluating such models in interaction was relatively straightforward since no additional integration work was required. However, data-driven approaches present a different situation: movements are typically generated in 3D space, which is more complicated to integrate into an interactive virtual agent. This is probably why previous systems did not evaluate learning-based models in an interactive setting. Users evaluating data-driven gesture generation models were typically asked to watch clips of ECAs' behaviors, rate the generated stimuli, or perform pairwise comparisons among ECAs \cite{Wolfert_Robinson_Belpaeme_2021}. In such approaches, humans are positioned as observers rather than as interactors, which is not as natural as the real-world scenarios faced by ECAs. In this study, we demonstrate how putting humans in interactors' place can be achieved, and we hope to inspire others to test their learning-based models in a similar manner - through interaction. 

The main contribution of this study is in evaluating interactively the use of deep-learning models to generate gestures. Our study also examines whether it is practical to use real-time interaction to measure the ECAs’ performance and compare the effect of their gesture behaviors on user perception. We also developed a testbed for future researchers to benchmark their models. The link to the video shows the experimental procedure used in our user study: \href{https://www.yaeh.io/research/hci/presentingbot}{www.yaeh.io/research/hci/presentingbot}

\section{Related Work}
In this section, we review both subjective and objective methods applied in previous studies to evaluate ECA with gestures. Then we discuss effective approaches to user studies, including common experimental settings, measurement tools, and evaluation dimensions, from which we can draw lessons for our own evaluation.

\subsection{Evaluation for ECAs with Gestures}
Most researchers use both objective and subjective methods to evaluate ECAs. On one hand, the researchers evaluate the performance of data-driven models objectively. On the other hand, researchers can understand how users perceive the system through subjective assessment, which is crucial for implementing systems in real-world applications.

Huang et al. conducted a subjective experiment to compare the human perception of a robot under four modal conditions 
\cite{Huang_Mutlu_2014}. The study results indicated that participants perceived higher levels of \textit{naturalness}, \textit{effectiveness}, and \textit{likability} when interacting with a robot that generated gestures based on a data-driven model than when interacting with a robot that communicated exclusively through speech.  


Levine et al. undertook a user study to evaluate a proposed HMM-based model. Participants were asked to perform side-by-side pairwise comparisons to determine which model produced more realistic gestures \cite{Levine2010}.

Yoon et al. evaluated the quality of gesture generation models by both objective and subjective methods. Objectively, the researchers compared the generated gestures with the ground truth in the original video. Subjectively, researchers adopted \textit{anthropomorphism} and \textit{likeability} from the Godspeed questionnaire \cite{Bartneck_2009} as measurements \cite{Yoon_Ko_Jang_Lee_Kim_Lee_2019}.

In 2020, a group of researchers launched the first gesture generation challenge \cite{kucherenko2021large}. This event provided a benchmark for practitioners in the gesture generation domain to compare their models by using a common dataset, visualization setup, and evaluation process. User studies were conducted in the challenge by watching video clips of different model results. In our study, we also focused on conducting user experiments, but through real-time interaction.

\subsection{Approaches to Conduct User Studies on ECAs with Gesture Functions}
\subsubsection{User Study through Interaction}
Real-time interactions with users are an effective means of assessing the performance of an ECA system or robot. In evaluating gesture-generation methods, it is also common to have participants rate the interaction experience. As an example, Salem et al. required participants to interact with a humanoid robot in both unimodal (speech only) and multimodal (speech with consistent or inconsistent gestures) conditions. The participants were then asked to rate a variety of aspects, including \textit{human-likeness}, \textit{likeability}, \textit{shared reality} with the robot, judgments of acceptance, and future contact intentions \cite{Salem_Eyssel_Rohlfing_Kopp_Joublin_2013}. Asly and Tapus examined personality traits (introversion and extroversion) in the design of gestural representations of robots, and corresponding user studies were conducted via real-time interactions \cite{Aly_Tapus_2013}. 

Nevertheless, our literature reviews found that real-time interaction studies tend to focus on rule-based methods.  In the case of measuring data-driven methods, user research is often conducted only in a one-way manner without the ability to communicate in real-time with the users. As an example, in Le and Pelachaud's perceptual experiment, participants were asked to rate the performance of gesture generation after viewing a video of an NAO robot narrating a fairy tale \cite{Le_Pelachaud}. GENEA 2020 Gesture-Generation Challenge \cite{kucherenko2021large} organizers provided participants with a user interface for measuring performance. This allowed them to rate their performance after viewing a video of the ECA, although the experiment did not involve interaction \cite{jonell2021hemvip}. A potential reason why data-driven methods were barely evaluated in real-time interactions could be the absence of appropriate testbeds. Due to the fact that data-driven methods are applied to generate gestures, the feedbacks from the ECAs are not fully predictable based on user input, which makes it more difficult to bind animations to ECAs. To tackle this limitation, Nagy et al. developed a modular framework in the Unity3D environment that enabled the ECA to communicate with data-driven models in real-time \cite{Nagy_2021}. Hence, it provided an infrastructure for future researchers to evaluate data-driven gesture generation models by adding modules. In this study, we adopted this framework and modified it to suit our scenario.

\subsubsection{Measurements to Evaluate Gesturing ECAs}
Gestures have a variety of effects on humans, including cognitive, emotional, and behavioral, as well as on their performance of tasks \cite{Saunderson_Nejat_2019}. However, because human communication is inherently multimodal, it is difficult to measure the contribution of gestures to human perception in isolation. To the best of our knowledge, there is still no unified questionnaire that can directly evaluate the quality of gestures in ECAs. Still, existing questionnaires can indirectly measure the role that gestures play in the system by assessing people's overall perception of ECA \cite{Fitrianie_2020}.

Researchers typically choose the tools for evaluation based on their research interests. Aly and Tagus' focus was on checking whether people can accurately identify robots that match their personality through non-verbal cues, and on testing whether gestures can improve expressiveness. In this study, experimenters used the Big 5 Inventory Test to distinguish between introverted and extroverted participants. Then experimenters adopted a questionnaire including 24 questions on a seven-point Likert scale to evaluate the gestures generation model, which is mapped by human personality traits. The survey covered user preferences, robot personality congruence, user engagement, robot expressiveness, robot gesture, voice synchronization, etc. \cite{Aly_Tapus_2013}. Salem et al. investigated the impact of robot gestures on humans' perceptions of anthropomorphism and likability. Researchers developed two sets of questionnaires to assess participants' perception of the robot and their performance on task-accomplishing interactions in the experiment. Perception-related questions included \textit{humanlikeness}, \textit{likeability}, \textit{shared reality} \cite{Echterhoff_Higgins_Levine_2009}, and \textit{future contact intentions}. To assess task performance, the authors used both objective and subjective ratings. The subjective indicator was a five-point Likert-scale question asking participants to rate their capacity to solve the challenge. The objective metric was the error rate of task completion \cite{Salem_Eyssel_Rohlfing_Kopp_Joublin_2013}.

Due to the lack of a standard evaluation scale, most studies have used self-developed questionnaires to assess gesture generation. Fitrianie et al. reviewed 81 papers relating to intelligent virtual agents, and found that more than 76\% of these studies employed measurement instruments that were only applied once \cite{Fitrianie_Bruijnes_Richards_Abdulrahman_Brinkman_2019}. Self-designed questionnaires do not have any guarantee of reliability. Additionally, constantly creating new measurement tools rather than reusing well-tested instruments makes it harder to replicate and compare experiments. Therefore, we applied validated and commonly-used measurements to our study to ensure its rigour and to facilitate comparison with other studies.

\subsubsection{Evaluating ECAs through Behavior Observation}
 Researchers can complement questionnaires with behavioral observations. Kooijmans et al. developed a software that can collect a variety of objective data in human-robot interactions, including sound, vision, person identification, motion, body contact, and robotic behavior \cite{Kooijmans_Kanda_Bartneck_Ishiguro_Hagita_2007}. Some other commonly utilized behavioral indicators include proxemics, postures \cite{Sanghvi_Castellano_Leite_Pereira_McOwan_Paiva_2011}, eye-gazes, etc. Bailenson et al. measured the social distance between a person and a virtual agent to determine co-presence in a virtual environment \cite{Bailenson_Guadagno_Aharoni_Dimov_Beall_Blascovich}. Nakano et al. used eye-gaze behavior as an objective indicator of a person's attention when talking to a virtual agent \cite{Nakano_Ishii_2010}. Behavioral indicators have been shown to be an effective complementary investigation method in conjunction with subjective questionnaires.

\section{System Design and Implementation}
The design and development of our ECA system were based on a realistic scenario of a virtual robot making a presentation. The virtual agent can present to participants six classical Roman monuments according to the participant's voice command. The script and the experiment system will be open-sourced.

\subsection{System Architecture}
\begin{figure*}[h]
  \centering
  \includegraphics[width=0.82\linewidth]{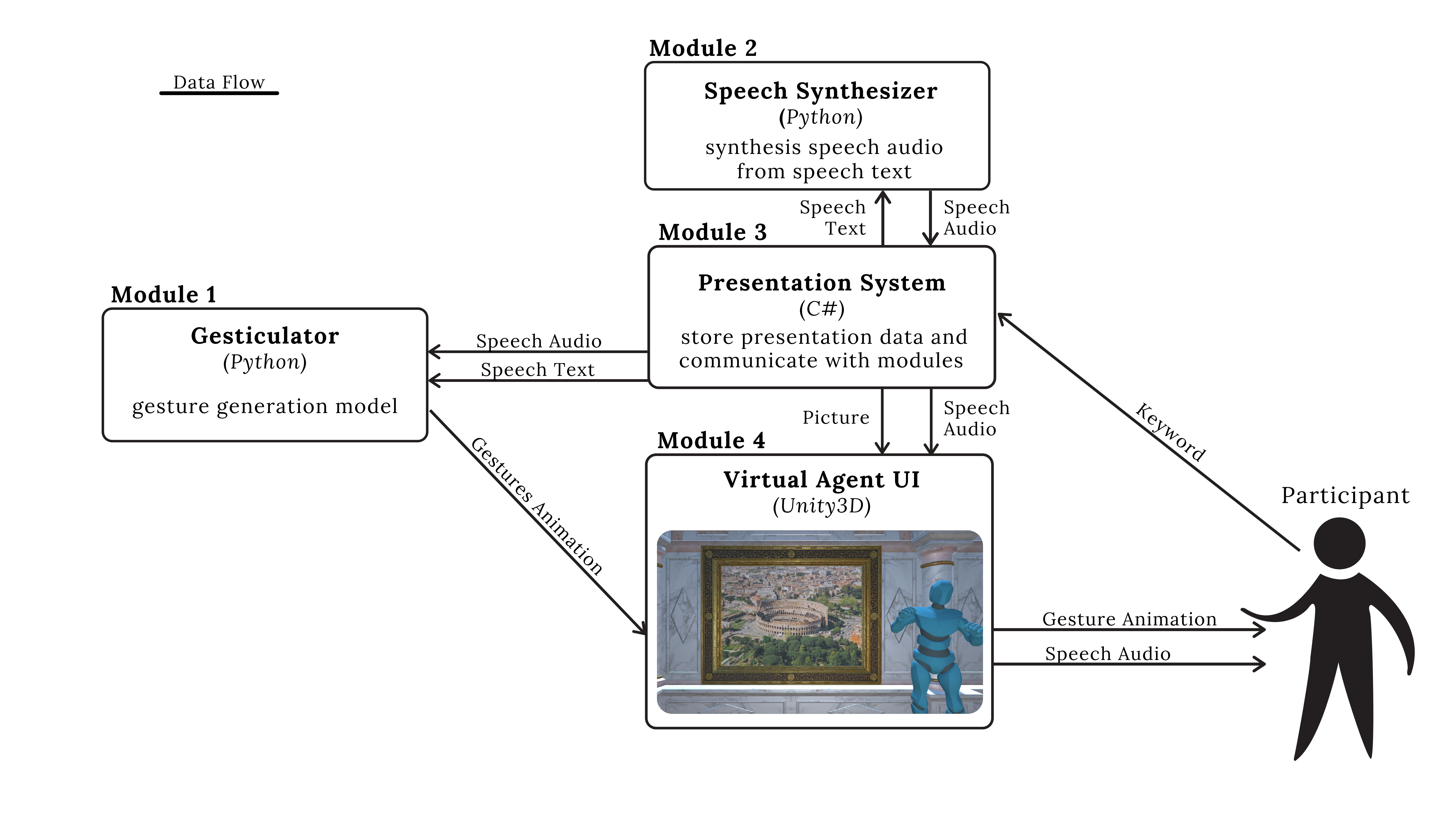}
  \caption{The system architecture.}
  \label{fig:architecture}
\end{figure*}

The system was based on the modular framework developed by Nagy et al \cite{Nagy_2021} \footnote{https://github.com/nagyrajmund/gesturebot}. The modified system architecture could be transplanted to any situation that involves a presenter, such as a lecturer or a museum guide. The system architecture consisted of four modules (see Figure \ref{fig:architecture}). Module 1 was the gesture generation model known as \textit{Gesticulator} developed by Kucherenko et al. \cite{Kucherenko_2020}. \textit{Gesticulator} is a data-driven method that can automate hand gestures based on both audio and text. Module 2 contained the speech synthesis model developed by Székely et al. \cite{szekely2019spontaneous}, which was used to generate the agent's audio output. This synthesizer was selected because it shared the same training dataset with our chosen gesture generation model, \textit{Gesticulator} \cite{Kucherenko_2020}. Module 3 was in charge of communication with other modules and contained presentation content. We developed a database in the Unity3D engine using C\#. The database allowed system users to customize the presentation's content by providing corresponding command words, pictures, audio, and speech text. Module 4 provided the virtual scene, which could be replaced by any customized 3D model. In this scene, the Gesticulator can communicate with the virtual robot through the ActiveMQ message broker. The current scene contains a virtual robot standing in front of a painting frame. The paintings are some famous Roman monuments. During interaction with participants, the painting content would swap like a presentation slider.  

\subsection{Modules Communication and Information Flow}
To avoid the system latency in response to participants' voice commands, we generated gestures in advance rather than in real-time. We input both presentation text and audio files into \textit{Gesticulator}, and \textit{Gesticulator} exported the generated gestures as joint angle rotations in CSV files. 

Before starting the experiment, we prepared four groups of presentation data. First, we wrote the presentation contents and saved them as text files. Second, we synthesized speech from the text using the speech synthesizer \cite{szekely2019spontaneous} and saved the outputs as audio files. Third, we selected pictures that would be shown in the presentation.  Fourth, we decided on the keywords that can invoke ECA's actions. We stored the four groups of resources and generated gestures in the database of Module 3. 

When the presentation system (Module 3) receives a keyword from a participant, it would send the related audio clip and image to the user interface (Module 4), while enabling the ECA to perform the corresponding gesture motion.

\section{User Study}

\subsection{Experiment Conditions and Hypotheses}
The purpose of this study was to evaluate the effects of co-speech gestures of the ECA system on user perception through interacting in real-time. Therefore, we identified two experimental conditions and referred to them as `Gesturing Condition' and `Idle Condition'. 

In both conditions, the virtual agent presents the Roman monuments selected by the participant's voice command. During the `Gesturing Condition', the virtual agent produces gestures while introducing the monument displayed on the painting. During `Idle Condition,' the virtual agent's hands merely hang naturally without gestures, accompanied by natural breathing and body microdynamics when presenting.

In order to collect each participant's subjective opinion towards both conditions, we applied a within-subjects experimental method, where each subject was exposed to two conditions. Instead of comparing different data-driven models, this study focused on examining the feasibility and approaches for evaluating data-driven gesture generation models in real-time, but the system design left room for future comparisons of multiple models.

In the study, each participant was assigned to both conditions. To exclude potential bias caused by ordering, we controlled the occurrence order of the two conditions. Fourteen participants were exposed to the gesturing condition first, followed by the idle condition; the remaining 14 participants were exposed to the opposite condition sequence. The robot body colors in both experimental conditions were set to blue and green, respectively, so that users could differentiate the two conditions. In order to avoid potential bias resulting from color preference, we counterbalanced the two colors and the two experimental conditions.

\begin{figure*}
  \centering
  \includegraphics[width=1.0\linewidth]{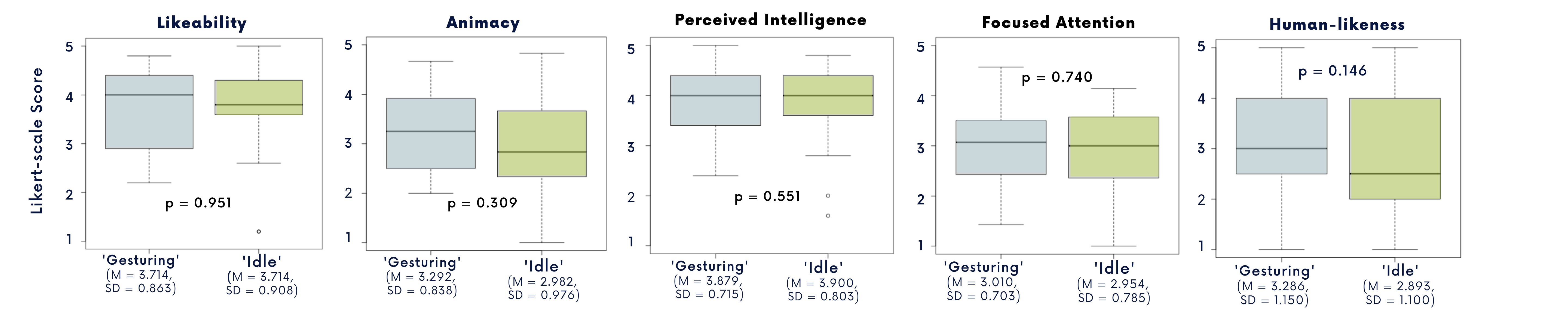}
  \caption{The questionnaire results on \textit{likeability}, \textit{animacy}, \textit{perceived intelligence}, \textit{focused attention} and \textit{human-likeness}}
  \label{fig:questionnaire result}
\end{figure*}
There was no significant difference in all those dimensions. The exact analytical values are shown in Figure \ref{fig:questionnaire result}.

We assumed that the ECA's hand gestures could improve human perception of the agent's realism and attract more attention during the interaction. To test this assumption, we propose two major hypotheses.

H1. The ECA with data-driven generated gestures is perceived better in \textit{likeability}, \textit{animacy}, \textit{perceived intelligence}, and \textit{human-likeness} than the ECA standing with idle posture.

H2. The ECA with data-driven generated gestures can attract more participant’s attention than the ECA standing with idle posture.

\subsection{Selection of Measuring Instruments}

To ensure our study's rigour, validity, and reproducibility, we chose to use a validated scale as our primary measurement tool. 

To examine hypothesis 1, we selected \textit{likeability}, \textit{animacy}, and \textit{perceived intelligence} dimensions from the Godspeed questionnaire \cite{Bartneck_2009} and asked a direct question regarding \textit{human-likeness}, which we took from the GENEA Challenge 2020 \cite{kucherenko2021large}. We asked participants to rate the interaction task subjectively in each condition. To evaluate hypothesis 2, both subjective and objective methods were adopted. Subjective assessment was measured using seven entries in the \textit{focused attention} dimension of User Engagement Scale (UES) \cite{OBrien2018}; Objective assessment was inferred from the gaze data collected by the eye tracker. It is believed that a longer gaze duration attracts more focused attention.

\subsection{Experiment Procedure}

Participants were exposed to two experimental conditions in succession. First, a selection menu guided the participant to select a keyword of interest. The participant would use a voice command to execute the selection. After receiving the command, the virtual robot would start presenting. Each speech segment lasted about 40 seconds. Each condition consisted of two selection interactions. After completing each experimental condition, the participant rated the interaction with each agent. From the start of the system, the user's eye gaze data was continuously recorded by a Tobii eye tracker at the bottom of the screen \cite{TobiiProLab}.


After the experiment, we interviewed each participant about  preferences and perceived differences between the two conditions.



At the end, the experimenter gave a short debriefing to each participant to explain the intention of the study.

\subsection{Participants}
A total of 28  participants (10 female, 18 male) were recruited by online and offline posters published in the university community (Facebook forums, Whatsapp groups, Telegram groups, advertisement boards and building doors on campus). After completing the experiment, participants were rewarded with a voucher from a local supermarket. 

Participants were diverse in age (18–36, Mean=24.39, SD=13.71) and nationality. To ensure the quality of the experiment, all participants were required to speak and listen to English at an excellent level.

\section{Results and Findings}
\subsection{Quantitative Results}
\subsubsection{Questionnaire}
The \textit{animacy} subscale consisted of six items ($\alpha$ = .90), the \textit{likeability} subscale consisted of five items ($\alpha$ = .91), the \textit{perceived intelligence} subscale consisted of five items ($\alpha$ = .85), and the \textit{focused attention} subscale consisted of seven items ($\alpha$ = .88). The measurements in all dimensions had good internal consistency.

We used a t-test to rule out a possible ordering bias to the data. Then, we performed the Shapiro-Wilk normality test on the mean values of each dimension. The results showed that \textit{animacy} and \textit{focused attention} could be considered to obey a normal distribution and could be tested by the paired t-test. In contrast, the distributions of the other three dimensions did not have normality and were suitable to be tested by a non-parametric test.

A paired samples t-test was performed to compare \textit{animacy} and \textit{focused attention} in `Gesturing' and `Idle' conditions.

An exact Wilcoxon-Pratt Signed-Rank Test was performed to compare \textit{likeability}, \textit{perceived intelligence} and \textit{human-likeness} in the `Gesturing' and `Idle' conditions.




\subsubsection{Eye-gazing Analysis}
\begin{figure}
  \centering
  \includegraphics[width=1.0\linewidth]{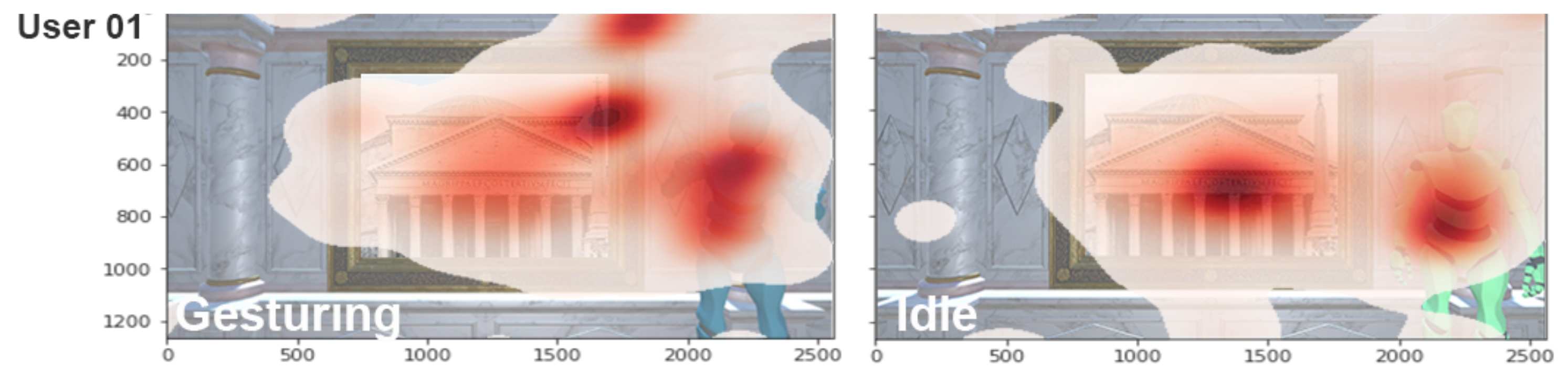}
  \caption{The heatmap shows gazing time (User 1 as example).}
  \label{fig:heatmap}
\end{figure}

\begin{figure}
  \centering
  \includegraphics[width=0.8\linewidth]{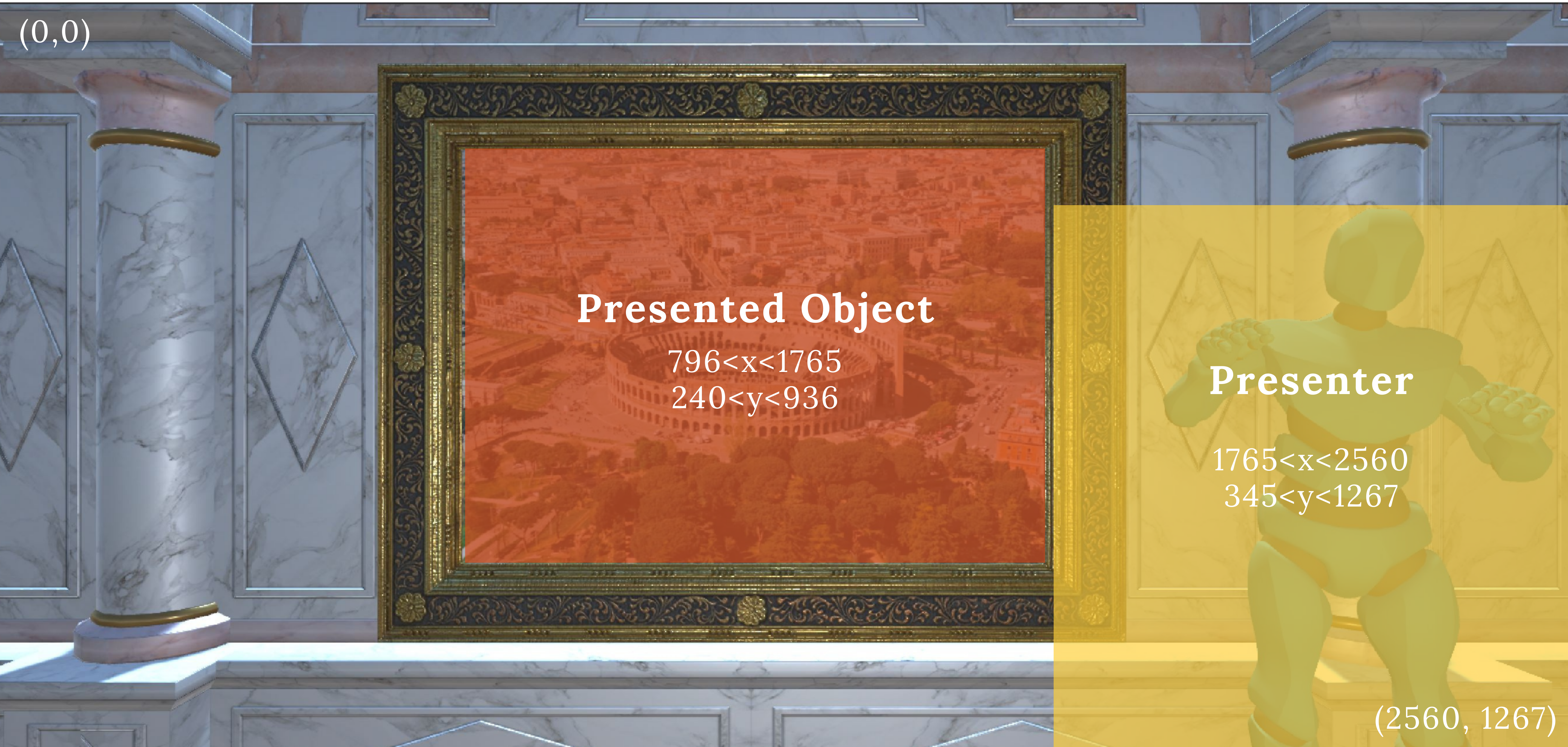}
  \caption{The division of presented object (painting) and the presenter (robot)}
  \label{fig:division}
\end{figure}

The original gaze data was recorded in the format of x, y coordinates for each frame. To establish an overall understanding of the subject's gaze behavior, we generated a separate heatmap (see Figure \ref{fig:heatmap}) for the two sets of coordinate data corresponding to each subject. Then, we divided the coordinate area corresponding to the screen into two parts, the presented object (painting) and the presenter (robot) (see 
Figure \ref{fig:division}). We obtain the ratio of gaze time corresponding to the two areas to the total interaction time and we subsequently express this proportion in terms of gazing time.

\begin{figure}[h]
  \centering
  \includegraphics[width=1\linewidth]{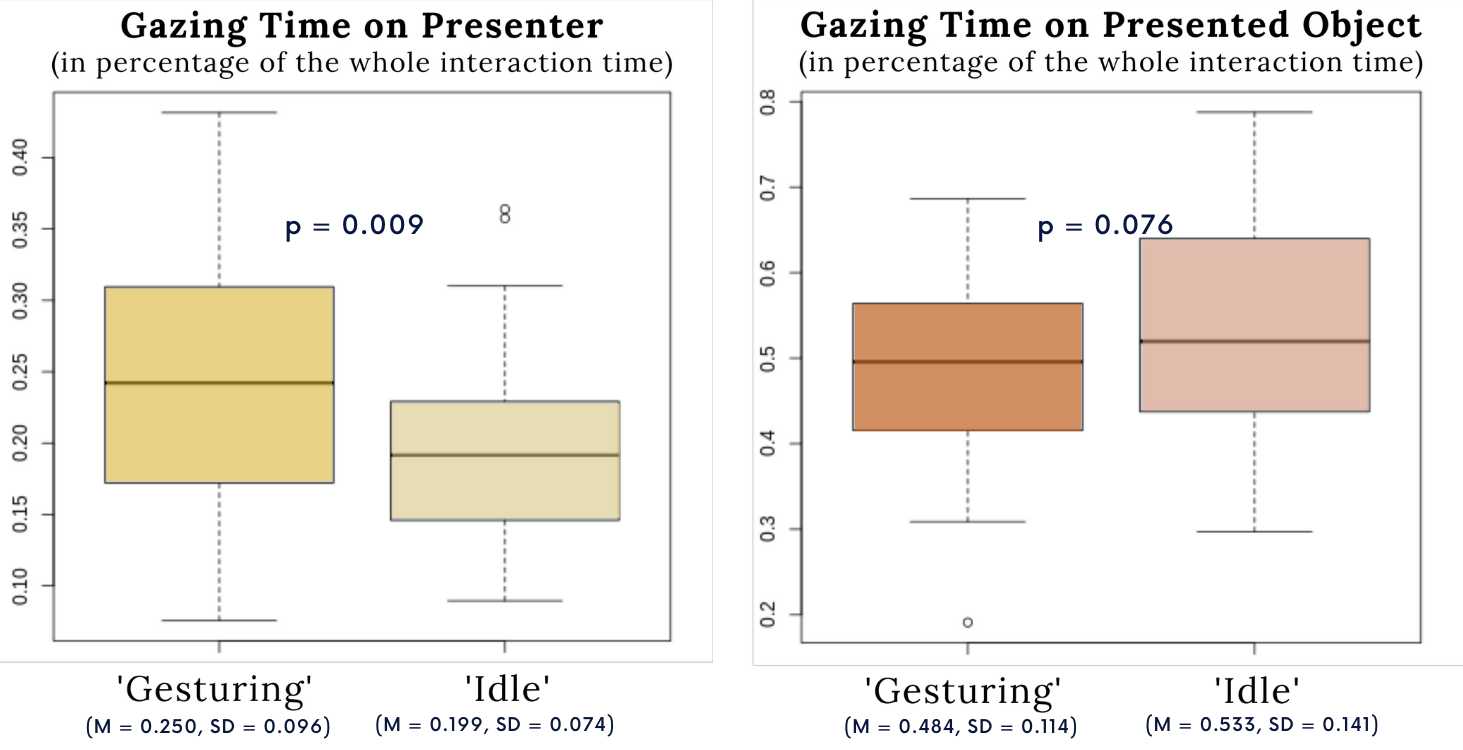}
  \caption{The boxplots of eye gazing time on presenter and presented object.}
  \label{fig:gazingTime}
\end{figure}

Figure \ref{fig:gazingTime} displays the distribution of results of eye gazing time on the virtual agent (presenter) and the painting (presented object) respectively.

A paired samples t-test was performed to compare eye gazing time on virtual agent (presenter) in `Gesturing' and `Idle' conditions. There was significant difference in eye gazing time on virtual agent (presenter)  between `Gesturing' (M = 0.250, SD = 0.096) and 'Idle' (M = 0.199, SD = 0.074); t(df) = 2.794, p = 0.009.

A paired samples t-test was performed to compare eye gazing time on the painting (presented object) in `Gesturing' and `Idle' conditions. There was no significant difference in eye gazing time on the painting (presented object)  between `Gesturing' (M = 0.484, SD = 0.114) and `Idle' (M = 0.533, SD = 0.141); t(df) = -1.844, p = 0.076.

\subsubsection{Findings}
\paragraph{Likeability}
As the result of our analysis in the \textit{Likeability} dimension shows, the participants did not prefer the virtual agent with gestures over the one with idle micro-movements. Figure  \ref{fig:questionnaire result} shows more participants rated the gesturing robot with low scores than the idle robot in \textit{likeability} dimension, though the mean score of the gesturing robot was still higher.  The result was consistent with the opinions received from the after-experiment interviews.

\paragraph{Human-likeness}
We found no significant difference between the two conditions. Both agents received extreme ratings in \textit{human-likeness} dimension.

\paragraph{Animacy}
In the analysis, the virtual agent with gestures received higher mean scores in \textit{animacy}. However, these differences did not reach statistical significance.

\paragraph{Perceived Intelligence}
\textit{Perceived intelligence} was the only dimension, where the `Idle' robot received slightly higher mean score than the  `Gesturing' robot. But the difference was not significant.

\paragraph{Focused Attention}
According to the questionnaire results, neither condition received more attention than the other. Nevertheless, eye gazing data analysis showed a significant difference when comparing gazing time on the agents and on the screen. Figure \ref{fig:questionnaire result} illustrates that participants spent more time on the presenter's body than presented object under `Gesturing' condition and vice versa. 
\subsection{Qualitative Result}
Two questions were asked in the after-experiment interviews:

Q1. Have you noticed the difference between two robots?

Q2. Which one do you prefer and why?

Regarding Q1, out of 28 participants, 24 mentioned color differences, 17 thought the two virtual agents' voice (pitch and tone) was different, and 15 called attention to the difference in hand movements.

Regarding the Q2, 13 participants preferred the virtual agent with gestures because it was more vivid, more natural, more expressive, friendlier, and that the movement could reduce boredom. Nine participants preferred the virtual agent without gestures because it looked calmer and less distracting. Six participants reported that it was hard to say which robot they preferred or that they liked neither of them. One of the participants explained that in this experiment, the virtual agent with gestures was too intensive and not natural enough. He argued that human-like gestures should only appear at specific times and not be continuously and uninterruptedly output. The virtual agent in the idle condition, on the other hand, was too stiff. One participant pointed out that the gesture animation appeared random and not anthropomorphic enough due to the virtual agent's inability to make iconic and deictic gestures.

As the result of our analysis in the \textit{likeability} dimension, the participants did not prefer the virtual agent with gestures over the one with idle micro-movements. Figure  \ref{fig:questionnaire result} showed more participants rated the gesturing robot with low scores than the idle robot in \textit{likeability} dimension, though the mean score of the gesturing robot was still higher.  The result was consistent with the opinions received from the after-experiment interviews.

We found no significant difference between the two conditions. For both agents, participants gave either very high or very low scores on the \textit{human-likeness} dimension.

In the analysis, the virtual agent with gestures received higher mean scores in \textit{animacy}. However, these differences did not reach statistical significance.

\textit{Perceived intelligence} was the only dimension, where the `Idle' robot received slightly higher mean score than the  `Gesturing' robot. But the difference was not significant.

According to the questionnaire results, neither condition received more attention than the other. Nevertheless, eye gazing data analysis showed a significant difference when comparing gazing time on the agents and on the screen. Figure \ref{fig:questionnaire result} illustrates that participants spent more time on the presenter's body than the presented object under `Gesturing' condition and vice versa.

\section{Discussion}

In this section, we discuss the results within the context of the proposed hypotheses and discuss the validity and usefulness of the suggested experimental methodology.

\subsection{H1: Improvement of \textit{likeability}, \textit{animacy}, \textit{perceived intelligence}, and \textit{human-likeness}}
Although the mean scores showed a slight indication that the ECA with data-driven generated gestures was perceived better in \textit{likeability},  \textit{animacy} and \textit{human-likeness} dimensions and that the ECA with idle posture received higher mean rate in the dimension of \textit{perceived intelligence}, the results were not statistically significant. Hence, we could not confirm H1. 

Other researchers, however, experienced different findings. For instance, in Salem et al.'s study, participants perceived the robot that gestured while it talked to be more likeable than the robot that only spoke verbally \cite{Salem_Eyssel_Rohlfing_Kopp_Joublin_2013}. Huang and Mutlu found that robots with gestures were more likeable than those without gestures \cite{Huang_Mutlu_2014}. Salem et al. \cite{Salem_Eyssel_Rohlfing_Kopp_Joublin_2013}  and Ishi et al. \cite{Ishi_Machiyashiki_Mikata_Ishiguro_2018} reported that a robot that could make co-speech gestures was considered more human-like than a standstill robot. 

By combining the participants' interview feedback with our analytical results, we speculate that the following factors may affect participants' ratings. First, the gesture-generation model we selected produced too many gestures that did not pause appropriately. In the presentation scenario, the presenters' excessive body movements may have distracted people from focusing on the presentation's content. Second, the dialogue system we adopted was not natural enough, and unexpected delays and lags could have affected the participants' \textit{likeability}. 

\subsection{H2: Increased attention}
Even though H2 was not supported by questionnaire results, it was confirmed by the result from eye gaze data. As the eye gazing results indicated, the gesturing agent attracted more attention to its body. On the other hand, the idle agent enabled participants to focus on the object being presented.

We were aware that using two different measurement tools could lead to contrasting results. The reason might be that when the user interface is complicated, the information obtained by the questionnaire is merely a general impression. We cannot know exactly which element on the interface is making the perceptual difference. In this case, the behavioral observation tool can be more sensitive for identifying the aspects that affect human perception.

\subsection{Experimental Methodology}

Our study also demonstrated the feasibility of evaluating data-driven motion generation models through real-time interactions. The experiment system worked smoothly and without unexpected interruption. From our experience in this study, we concluded three advantages to conducting evaluation practices through real-time interaction rather than watching video clips.

First, evaluating through real-time interaction is more natural than watching video clips because it simulates real-world communication with humans. Participants notice more details when they interact with agents. For example, the way a virtual agent reacts to some human behaviors can be perceived as an indication of intelligence.

Second, the interaction process can be designed to adapt to the actual use scenarios. Evaluating specific use scenarios could lead to varied results. For example, \textit{likeability} could be highly dependant on the use scenario. A virtual agent that is designed to be extremely vivacious might be perceived as likeable when presented as a dance teacher but less likeable as a technical expert.

Third, involving real-time interaction in evaluation practices allows for more possibilities to employ behavioral observation methods. By analyzing human responses to the system during the interaction, such as posture and eye gazing, experimenters could obtain richer results.

\section{Limitations and Future Work}
The current study has several limitations. The small number of participants meant that we could not get a sufficiently diverse sample. Our literature review shows that different cultures and sexes perceive gestures differently. In light of this, we speculate we will obtain richer results if we can recruit more participants. If there is a chance for a follow-up study, we will launch the existing test system online to reach more participants.

This study evaluated only one data-driven model for its impact on user perception due to time constraints. The existing system design permits the integration of multiple virtual robots and gesture models, enabling future comparisons between the models.

The experiment was conducted in a desktop environment, where users could not interact in a fully immersive way with the virtual robot. As indicated by our findings, objective user behavior data can be beneficial for supplementary and productive outcomes. For further study, it could be useful to enable the system to collect multiple types of data. Virtual reality environments may be able to immerse users while also collecting objective data on user behavior (e.g., social distance).

\section{Conclusion}

In this paper, we proposed an approach for evaluating data-driven gesture generation models in an interaction. To test the framework we compared two conditions: one with idle gestures and one with gestures generated by a machine-learning model. 
Even though the questionnaire results did not indicate a significant difference in all the dimensions, the analysis of eye gazing data has supported the second hypothesis — \textbf{the ECA with data-driven generated gestures can attract more participants' attention in certain areas than the ECA standing with idle posture}.

The finding can be used as an empirical basis for the design of ECA applications in presentation scenarios. For example, a math teacher agent would require students to pay more attention to the slides than to the agent, so gestures would be less of a factor. In contrast, an NPC in a game that tells the history of a particular world may benefit from incorporating gestures to attract the player's attention and make the character more expressive.

In addition to subjective questionnaires, objective data can provide valuable insight. Our study discussed and validated the use of objective behavioral data to complement structured questionnaire data. Our study used gaze time to infer \textit{focused attention} — longer gaze times are considered to indicate greater focus \cite{Frischen_Bayliss_Tipper_2007}. 

Our experimental setup serves as an example of how real-time evaluations of data-driven gesture models can be performed efficiently and effectively. The results of our study suggest that conducting user studies with behavioral observation methods, such as eye-tracking, can be beneficial.

\begin{acks}
The authors are grateful to Rajmund Nagy for helping to set up the Gesture Bot system, as well as to Éva Székely for kindly providing her speech synthesis model. We are also immensely grateful to Jim Royal for proofreading the paper.
\end{acks}

\newpage

\balance

\bibliographystyle{ACM-Reference-Format}
\bibliography{sample-base}


\begin{thebibliography}{40}


\ifx \showCODEN    \undefined \def \showCODEN     #1{\unskip}     \fi
\ifx \showDOI      \undefined \def \showDOI       #1{#1}\fi
\ifx \showISBNx    \undefined \def \showISBNx     #1{\unskip}     \fi
\ifx \showISBNxiii \undefined \def \showISBNxiii  #1{\unskip}     \fi
\ifx \showISSN     \undefined \def \showISSN      #1{\unskip}     \fi
\ifx \showLCCN     \undefined \def \showLCCN      #1{\unskip}     \fi
\ifx \shownote     \undefined \def \shownote      #1{#1}          \fi
\ifx \showarticletitle \undefined \def \showarticletitle #1{#1}   \fi
\ifx \showURL      \undefined \def \showURL       {\relax}        \fi
\providecommand\bibfield[2]{#2}
\providecommand\bibinfo[2]{#2}
\providecommand\natexlab[1]{#1}
\providecommand\showeprint[2][]{arXiv:#2}

\bibitem[Alexanderson et~al\mbox{.}(2020)]%
        {Alexanderson_Henter_Kucherenko_Beskow_2020}
\bibfield{author}{\bibinfo{person}{Simon Alexanderson},
  \bibinfo{person}{Gustav~Eje Henter}, \bibinfo{person}{Taras Kucherenko},
  {and} \bibinfo{person}{Jonas Beskow}.} \bibinfo{year}{2020}\natexlab{}.
\newblock \showarticletitle{Style-Controllable Speech-Driven Gesture Synthesis
  Using Normalising Flows}. In \bibinfo{booktitle}{\emph{Computer Graphics
  Forum}}, Vol.~\bibinfo{volume}{39}. Wiley Online Library,
  \bibinfo{pages}{487--496}.
\newblock


\bibitem[Aly and Tapus(2013)]%
        {Aly_Tapus_2013}
\bibfield{author}{\bibinfo{person}{Amir Aly} {and} \bibinfo{person}{Adriana
  Tapus}.} \bibinfo{year}{2013}\natexlab{}.
\newblock \showarticletitle{A model for synthesizing a combined verbal and
  nonverbal behavior based on personality traits in human-robot interaction}.
  In \bibinfo{booktitle}{\emph{2013 8th ACM/IEEE International Conference on
  Human-Robot Interaction (HRI)}}. IEEE, \bibinfo{pages}{325--332}.
\newblock


\bibitem[Bailenson et~al\mbox{.}(2004)]%
        {Bailenson_Guadagno_Aharoni_Dimov_Beall_Blascovich}
\bibfield{author}{\bibinfo{person}{Jeremy~N Bailenson}, \bibinfo{person}{Eyal
  Aharoni}, \bibinfo{person}{Andrew~C Beall}, \bibinfo{person}{Rosanna~E
  Guadagno}, \bibinfo{person}{Aleksandar Dimov}, {and} \bibinfo{person}{Jim
  Blascovich}.} \bibinfo{year}{2004}\natexlab{}.
\newblock \showarticletitle{Comparing behavioral and self-report measures of
  embodied agents’ social presence in immersive virtual environments}. In
  \bibinfo{booktitle}{\emph{Proceedings of the 7th Annual International
  Workshop on PRESENCE}}. IEEE, \bibinfo{pages}{1864--1105}.
\newblock


\bibitem[Bartneck et~al\mbox{.}(2009)]%
        {Bartneck_2009}
\bibfield{author}{\bibinfo{person}{Christoph Bartneck}, \bibinfo{person}{Dana
  Kuli{\'c}}, \bibinfo{person}{Elizabeth Croft}, {and} \bibinfo{person}{Susana
  Zoghbi}.} \bibinfo{year}{2009}\natexlab{}.
\newblock \showarticletitle{Measurement instruments for the anthropomorphism,
  animacy, likeability, perceived intelligence, and perceived safety of
  robots}.
\newblock \bibinfo{journal}{\emph{International journal of social robotics}}
  \bibinfo{volume}{1}, \bibinfo{number}{1} (\bibinfo{year}{2009}),
  \bibinfo{pages}{71--81}.
\newblock


\bibitem[Cassell et~al\mbox{.}(2000)]%
        {Cassell2001}
\bibfield{author}{\bibinfo{person}{Justine Cassell}, \bibinfo{person}{Tim
  Bickmore}, \bibinfo{person}{Lee Campbell}, \bibinfo{person}{Hannes
  Vilhjalmsson}, {and} \bibinfo{person}{Hao Yan}.}
  \bibinfo{year}{2000}\natexlab{}.
\newblock \showarticletitle{Human conversation as a system framework: Designing
  embodied conversational agents}.
\newblock \bibinfo{journal}{\emph{Embodied conversational agents}}
  (\bibinfo{year}{2000}), \bibinfo{pages}{29--63}.
\newblock


\bibitem[Cassell et~al\mbox{.}(1994)]%
        {Cassell_Pelachaud_1994}
\bibfield{author}{\bibinfo{person}{Justine Cassell}, \bibinfo{person}{Catherine
  Pelachaud}, \bibinfo{person}{Norman Badler}, \bibinfo{person}{Mark Steedman},
  \bibinfo{person}{Brett Achorn}, \bibinfo{person}{Tripp Becket},
  \bibinfo{person}{Brett Douville}, \bibinfo{person}{Scott Prevost}, {and}
  \bibinfo{person}{Matthew Stone}.} \bibinfo{year}{1994}\natexlab{}.
\newblock \showarticletitle{Animated conversation: rule-based generation of
  facial expression, gesture \& spoken intonation for multiple conversational
  agents}. In \bibinfo{booktitle}{\emph{Proceedings of the 21st annual
  conference on Computer graphics and interactive techniques}}.
  \bibinfo{pages}{413--420}.
\newblock


\bibitem[Cassell and Vilhj{\'a}lmsson(1999)]%
        {Cassell_Vilhja}
\bibfield{author}{\bibinfo{person}{Justine Cassell} {and}
  \bibinfo{person}{Hannes Vilhj{\'a}lmsson}.} \bibinfo{year}{1999}\natexlab{}.
\newblock \showarticletitle{Fully embodied conversational avatars: Making
  communicative behaviors autonomous}.
\newblock \bibinfo{journal}{\emph{Autonomous agents and multi-agent systems}}
  \bibinfo{volume}{2}, \bibinfo{number}{1} (\bibinfo{year}{1999}),
  \bibinfo{pages}{45--64}.
\newblock


\bibitem[Chiu and Marsella(2011)]%
        {Chiu_Marsella_2011}
\bibfield{author}{\bibinfo{person}{Chung-Cheng Chiu} {and}
  \bibinfo{person}{Stacy Marsella}.} \bibinfo{year}{2011}\natexlab{}.
\newblock \showarticletitle{How to train your avatar: A data driven approach to
  gesture generation}. In \bibinfo{booktitle}{\emph{International Workshop on
  Intelligent Virtual Agents}}. Springer, \bibinfo{pages}{127--140}.
\newblock


\bibitem[Chong et~al\mbox{.}(2018)]%
        {Chong_2018}
\bibfield{author}{\bibinfo{person}{Eunji Chong}, \bibinfo{person}{Nataniel
  Ruiz}, \bibinfo{person}{Yongxin Wang}, \bibinfo{person}{Yun Zhang},
  \bibinfo{person}{Agata Rozga}, {and} \bibinfo{person}{James~M Rehg}.}
  \bibinfo{year}{2018}\natexlab{}.
\newblock \showarticletitle{Connecting gaze, scene, and attention: Generalized
  attention estimation via joint modeling of gaze and scene saliency}. In
  \bibinfo{booktitle}{\emph{Proceedings of the European conference on computer
  vision (ECCV)}}. \bibinfo{pages}{383--398}.
\newblock


\bibitem[Echterhoff et~al\mbox{.}(2009)]%
        {Echterhoff_Higgins_Levine_2009}
\bibfield{author}{\bibinfo{person}{Gerald Echterhoff}, \bibinfo{person}{E~Tory
  Higgins}, {and} \bibinfo{person}{John~M Levine}.}
  \bibinfo{year}{2009}\natexlab{}.
\newblock \showarticletitle{Shared reality: Experiencing commonality with
  others' inner states about the world}.
\newblock \bibinfo{journal}{\emph{Perspectives on Psychological Science}}
  \bibinfo{volume}{4}, \bibinfo{number}{5} (\bibinfo{year}{2009}),
  \bibinfo{pages}{496--521}.
\newblock


\bibitem[Fabri et~al\mbox{.}(2002)]%
        {M_expressiveagents}
\bibfield{author}{\bibinfo{person}{Marc Fabri}, \bibinfo{person}{DJ Moore},
  {and} \bibinfo{person}{DJ Hobbs}.} \bibinfo{year}{2002}\natexlab{}.
\newblock \showarticletitle{Expressive agents: Non-verbal communication in
  collaborative virtual environments}.
\newblock \bibinfo{journal}{\emph{Proceedings of Autonomous Agents and
  Multi-Agent Systems (Embodied Conversational Agents)}}
  (\bibinfo{year}{2002}).
\newblock


\bibitem[Feine et~al\mbox{.}(2019)]%
        {Feine_Gnewuch_Morana_Maedche_2019}
\bibfield{author}{\bibinfo{person}{Jasper Feine}, \bibinfo{person}{Ulrich
  Gnewuch}, \bibinfo{person}{Stefan Morana}, {and} \bibinfo{person}{Alexander
  Maedche}.} \bibinfo{year}{2019}\natexlab{}.
\newblock \showarticletitle{A taxonomy of social cues for conversational
  agents}.
\newblock \bibinfo{journal}{\emph{International Journal of Human-Computer
  Studies}}  \bibinfo{volume}{132} (\bibinfo{year}{2019}),
  \bibinfo{pages}{138--161}.
\newblock


\bibitem[Fitrianie et~al\mbox{.}(2019)]%
        {Fitrianie_Bruijnes_Richards_Abdulrahman_Brinkman_2019}
\bibfield{author}{\bibinfo{person}{Siska Fitrianie}, \bibinfo{person}{Merijn
  Bruijnes}, \bibinfo{person}{Deborah Richards}, \bibinfo{person}{Amal
  Abdulrahman}, {and} \bibinfo{person}{Willem-Paul Brinkman}.}
  \bibinfo{year}{2019}\natexlab{}.
\newblock \showarticletitle{What are We Measuring Anyway? -A Literature Survey
  of Questionnaires Used in Studies Reported in the Intelligent Virtual Agent
  Conferences}. In \bibinfo{booktitle}{\emph{Proceedings of the 19th ACM
  International Conference on Intelligent Virtual Agents}}.
  \bibinfo{pages}{159--161}.
\newblock


\bibitem[Fitrianie et~al\mbox{.}(2020)]%
        {Fitrianie_2020}
\bibfield{author}{\bibinfo{person}{Siska Fitrianie}, \bibinfo{person}{Merijn
  Bruijnes}, \bibinfo{person}{Deborah Richards}, \bibinfo{person}{Andrea
  B{\"o}nsch}, {and} \bibinfo{person}{Willem-Paul Brinkman}.}
  \bibinfo{year}{2020}\natexlab{}.
\newblock \showarticletitle{The 19 unifying questionnaire constructs of
  artificial social agents: An iva community analysis}. In
  \bibinfo{booktitle}{\emph{Proceedings of the 20th ACM International
  Conference on Intelligent Virtual Agents}}. \bibinfo{pages}{1--8}.
\newblock


\bibitem[Frischen et~al\mbox{.}(2007)]%
        {Frischen_Bayliss_Tipper_2007}
\bibfield{author}{\bibinfo{person}{Alexandra Frischen},
  \bibinfo{person}{Andrew~P Bayliss}, {and} \bibinfo{person}{Steven~P Tipper}.}
  \bibinfo{year}{2007}\natexlab{}.
\newblock \showarticletitle{Gaze cueing of attention: visual attention, social
  cognition, and individual differences.}
\newblock \bibinfo{journal}{\emph{Psychological bulletin}}
  \bibinfo{volume}{133}, \bibinfo{number}{4} (\bibinfo{year}{2007}),
  \bibinfo{pages}{694}.
\newblock


\bibitem[Huang and Mutlu(2013)]%
        {Huang_Mutlu_2013}
\bibfield{author}{\bibinfo{person}{Chien-Ming Huang} {and}
  \bibinfo{person}{Bilge Mutlu}.} \bibinfo{year}{2013}\natexlab{}.
\newblock \showarticletitle{Modeling and Evaluating Narrative Gestures for
  Humanlike Robots.}. In \bibinfo{booktitle}{\emph{Robotics: Science and
  Systems}}. \bibinfo{pages}{57--64}.
\newblock


\bibitem[Huang and Mutlu(2014)]%
        {Huang_Mutlu_2014}
\bibfield{author}{\bibinfo{person}{Chien-Ming Huang} {and}
  \bibinfo{person}{Bilge Mutlu}.} \bibinfo{year}{2014}\natexlab{}.
\newblock \showarticletitle{Learning-based modeling of multimodal behaviors for
  humanlike robots}. In \bibinfo{booktitle}{\emph{2014 9th ACM/IEEE
  International Conference on Human-Robot Interaction (HRI)}}. IEEE,
  \bibinfo{pages}{57--64}.
\newblock


\bibitem[Ishi et~al\mbox{.}(2018)]%
        {Ishi_Machiyashiki_Mikata_Ishiguro_2018}
\bibfield{author}{\bibinfo{person}{Carlos~T Ishi}, \bibinfo{person}{Daichi
  Machiyashiki}, \bibinfo{person}{Ryusuke Mikata}, {and}
  \bibinfo{person}{Hiroshi Ishiguro}.} \bibinfo{year}{2018}\natexlab{}.
\newblock \showarticletitle{A speech-driven hand gesture generation method and
  evaluation in android robots}.
\newblock \bibinfo{journal}{\emph{IEEE Robotics and Automation Letters}}
  \bibinfo{volume}{3}, \bibinfo{number}{4} (\bibinfo{year}{2018}),
  \bibinfo{pages}{3757--3764}.
\newblock


\bibitem[Jonell et~al\mbox{.}(2021)]%
        {jonell2021hemvip}
\bibfield{author}{\bibinfo{person}{Patrik Jonell}, \bibinfo{person}{Youngwoo
  Yoon}, \bibinfo{person}{Pieter Wolfert}, \bibinfo{person}{Taras Kucherenko},
  {and} \bibinfo{person}{Gustav~Eje Henter}.} \bibinfo{year}{2021}\natexlab{}.
\newblock \showarticletitle{HEMVIP: Human evaluation of multiple videos in
  parallel}. In \bibinfo{booktitle}{\emph{Proceedings of the 2021 International
  Conference on Multimodal Interaction}}. \bibinfo{pages}{707--711}.
\newblock


\bibitem[Knapp et~al\mbox{.}(2013)]%
        {Knapp_Hall_Horgan_2012}
\bibfield{author}{\bibinfo{person}{Mark~L Knapp}, \bibinfo{person}{Judith~A
  Hall}, {and} \bibinfo{person}{Terrence~G Horgan}.}
  \bibinfo{year}{2013}\natexlab{}.
\newblock \bibinfo{booktitle}{\emph{Nonverbal communication in human
  interaction}}.
\newblock \bibinfo{publisher}{Cengage Learning}.
\newblock


\bibitem[Kooijmans et~al\mbox{.}(2007)]%
        {Kooijmans_Kanda_Bartneck_Ishiguro_Hagita_2007}
\bibfield{author}{\bibinfo{person}{Tijn Kooijmans}, \bibinfo{person}{Takayuki
  Kanda}, \bibinfo{person}{Christoph Bartneck}, \bibinfo{person}{Hiroshi
  Ishiguro}, {and} \bibinfo{person}{Norihiro Hagita}.}
  \bibinfo{year}{2007}\natexlab{}.
\newblock \showarticletitle{Accelerating robot development through integral
  analysis of human--robot interaction}.
\newblock \bibinfo{journal}{\emph{IEEE Transactions on Robotics}}
  \bibinfo{volume}{23}, \bibinfo{number}{5} (\bibinfo{year}{2007}),
  \bibinfo{pages}{1001--1012}.
\newblock


\bibitem[Kucherenko et~al\mbox{.}(2021a)]%
        {kucherenko2021moving}
\bibfield{author}{\bibinfo{person}{Taras Kucherenko}, \bibinfo{person}{Dai
  Hasegawa}, \bibinfo{person}{Naoshi Kaneko}, \bibinfo{person}{Gustav~Eje
  Henter}, {and} \bibinfo{person}{Hedvig Kjellstr{\"o}m}.}
  \bibinfo{year}{2021}\natexlab{a}.
\newblock \showarticletitle{Moving fast and slow: Analysis of representations
  and post-processing in speech-driven automatic gesture generation}.
\newblock \bibinfo{journal}{\emph{International Journal of Human--Computer
  Interaction}} (\bibinfo{year}{2021}), \bibinfo{pages}{1--17}.
\newblock


\bibitem[Kucherenko et~al\mbox{.}(2020)]%
        {Kucherenko_2020}
\bibfield{author}{\bibinfo{person}{Taras Kucherenko}, \bibinfo{person}{Patrik
  Jonell}, \bibinfo{person}{Sanne van Waveren}, \bibinfo{person}{Gustav~Eje
  Henter}, \bibinfo{person}{Simon Alexandersson}, \bibinfo{person}{Iolanda
  Leite}, {and} \bibinfo{person}{Hedvig Kjellstr{\"o}m}.}
  \bibinfo{year}{2020}\natexlab{}.
\newblock \showarticletitle{Gesticulator: A framework for semantically-aware
  speech-driven gesture generation}. In \bibinfo{booktitle}{\emph{Proceedings
  of the 2020 International Conference on Multimodal Interaction}}.
  \bibinfo{pages}{242--250}.
\newblock


\bibitem[Kucherenko et~al\mbox{.}(2021b)]%
        {kucherenko2021large}
\bibfield{author}{\bibinfo{person}{Taras Kucherenko}, \bibinfo{person}{Patrik
  Jonell}, \bibinfo{person}{Youngwoo Yoon}, \bibinfo{person}{Pieter Wolfert},
  {and} \bibinfo{person}{Gustav~Eje Henter}.} \bibinfo{year}{2021}\natexlab{b}.
\newblock \showarticletitle{A large, crowdsourced evaluation of gesture
  generation systems on common data: The GENEA Challenge 2020}. In
  \bibinfo{booktitle}{\emph{26th International Conference on Intelligent User
  Interfaces}}. \bibinfo{pages}{11--21}.
\newblock


\bibitem[Le and Pelachaud(2012)]%
        {Le_Pelachaud}
\bibfield{author}{\bibinfo{person}{Quoc~Anh Le} {and}
  \bibinfo{person}{Catherine Pelachaud}.} \bibinfo{year}{2012}\natexlab{}.
\newblock \showarticletitle{Evaluating an expressive gesture model for a
  humanoid robot: Experimental results}. In \bibinfo{booktitle}{\emph{Submitted
  to 8th ACM/IEEE International Conference on Human-Robot Interaction}}.
\newblock


\bibitem[Levine et~al\mbox{.}(2010)]%
        {Levine2010}
\bibfield{author}{\bibinfo{person}{Sergey Levine}, \bibinfo{person}{Philipp
  Kr{\"a}henb{\"u}hl}, \bibinfo{person}{Sebastian Thrun}, {and}
  \bibinfo{person}{Vladlen Koltun}.} \bibinfo{year}{2010}\natexlab{}.
\newblock \showarticletitle{Gesture controllers}.
\newblock In \bibinfo{booktitle}{\emph{ACM SIGGRAPH 2010 papers}}.
  \bibinfo{pages}{1--11}.
\newblock


\bibitem[Marmpena et~al\mbox{.}(2020)]%
        {Marmpena_Garcia_Lim_2020}
\bibfield{author}{\bibinfo{person}{Mina Marmpena}, \bibinfo{person}{Fernando
  Garcia}, {and} \bibinfo{person}{Angelica Lim}.}
  \bibinfo{year}{2020}\natexlab{}.
\newblock \showarticletitle{Generating robotic emotional body language of
  targeted valence and arousal with conditional variational autoencoders}. In
  \bibinfo{booktitle}{\emph{Companion of the 2020 ACM/IEEE international
  conference on human-robot interaction}}. \bibinfo{pages}{357--359}.
\newblock


\bibitem[Nagy et~al\mbox{.}(2021)]%
        {Nagy_2021}
\bibfield{author}{\bibinfo{person}{Rajmund Nagy}, \bibinfo{person}{Taras
  Kucherenko}, \bibinfo{person}{Birger Moell}, \bibinfo{person}{Andr\'{e}
  Pereira}, \bibinfo{person}{Hedvig Kjellstr\"{o}m}, {and}
  \bibinfo{person}{Ulysses Bernardet}.} \bibinfo{year}{2021}\natexlab{}.
\newblock \showarticletitle{A Framework for Integrating Gesture Generation
  Models into Interactive Conversational Agents}. In
  \bibinfo{booktitle}{\emph{Proceedings of the 20th International Conference on
  Autonomous Agents and MultiAgent Systems}} \emph{(\bibinfo{series}{AAMAS
  '21})}. \bibinfo{pages}{1779–1781}.
\newblock


\bibitem[Nakano and Ishii(2010)]%
        {Nakano_Ishii_2010}
\bibfield{author}{\bibinfo{person}{Yukiko~I Nakano} {and} \bibinfo{person}{Ryo
  Ishii}.} \bibinfo{year}{2010}\natexlab{}.
\newblock \showarticletitle{Estimating user's engagement from eye-gaze
  behaviors in human-agent conversations}. In
  \bibinfo{booktitle}{\emph{Proceedings of the 15th international conference on
  Intelligent user interfaces}}. \bibinfo{pages}{139--148}.
\newblock


\bibitem[O’Brien et~al\mbox{.}(2018)]%
        {OBrien2018}
\bibfield{author}{\bibinfo{person}{Heather~L O’Brien}, \bibinfo{person}{Paul
  Cairns}, {and} \bibinfo{person}{Mark Hall}.} \bibinfo{year}{2018}\natexlab{}.
\newblock \showarticletitle{A practical approach to measuring user engagement
  with the refined user engagement scale (UES) and new UES short form}.
\newblock \bibinfo{journal}{\emph{International Journal of Human-Computer
  Studies}}  \bibinfo{volume}{112} (\bibinfo{year}{2018}),
  \bibinfo{pages}{28--39}.
\newblock


\bibitem[Pelachaud(2015)]%
        {pelachaud2015greta}
\bibfield{author}{\bibinfo{person}{Catherine Pelachaud}.}
  \bibinfo{year}{2015}\natexlab{}.
\newblock \showarticletitle{Greta: an interactive expressive embodied
  conversational agent}. In \bibinfo{booktitle}{\emph{Proceedings of the 2015
  International Conference on Autonomous Agents and Multiagent Systems}}.
  \bibinfo{pages}{5--5}.
\newblock


\bibitem[Ravenet et~al\mbox{.}(2018)]%
        {Ravenet_Pelachaud_Clavel_Marsella_2018}
\bibfield{author}{\bibinfo{person}{Brian Ravenet}, \bibinfo{person}{Catherine
  Pelachaud}, \bibinfo{person}{Chlo{\'e} Clavel}, {and} \bibinfo{person}{Stacy
  Marsella}.} \bibinfo{year}{2018}\natexlab{}.
\newblock \showarticletitle{Automating the production of communicative gestures
  in embodied characters}.
\newblock \bibinfo{journal}{\emph{Frontiers in psychology}}
  \bibinfo{volume}{9} (\bibinfo{year}{2018}), \bibinfo{pages}{1144}.
\newblock


\bibitem[Salem et~al\mbox{.}(2013)]%
        {Salem_Eyssel_Rohlfing_Kopp_Joublin_2013}
\bibfield{author}{\bibinfo{person}{Maha Salem}, \bibinfo{person}{Friederike
  Eyssel}, \bibinfo{person}{Katharina Rohlfing}, \bibinfo{person}{Stefan Kopp},
  {and} \bibinfo{person}{Frank Joublin}.} \bibinfo{year}{2013}\natexlab{}.
\newblock \showarticletitle{To err is human (-like): Effects of robot gesture
  on perceived anthropomorphism and likability}.
\newblock \bibinfo{journal}{\emph{International Journal of Social Robotics}}
  \bibinfo{volume}{5}, \bibinfo{number}{3} (\bibinfo{year}{2013}),
  \bibinfo{pages}{313--323}.
\newblock


\bibitem[Sanghvi et~al\mbox{.}(2011)]%
        {Sanghvi_Castellano_Leite_Pereira_McOwan_Paiva_2011}
\bibfield{author}{\bibinfo{person}{Jyotirmay Sanghvi}, \bibinfo{person}{Ginevra
  Castellano}, \bibinfo{person}{Iolanda Leite}, \bibinfo{person}{Andr{\'e}
  Pereira}, \bibinfo{person}{Peter~W McOwan}, {and} \bibinfo{person}{Ana
  Paiva}.} \bibinfo{year}{2011}\natexlab{}.
\newblock \showarticletitle{Automatic analysis of affective postures and body
  motion to detect engagement with a game companion}. In
  \bibinfo{booktitle}{\emph{Proceedings of the 6th international conference on
  Human-robot interaction}}. \bibinfo{pages}{305--312}.
\newblock


\bibitem[Saunderson and Nejat(2019)]%
        {Saunderson_Nejat_2019}
\bibfield{author}{\bibinfo{person}{Shane Saunderson} {and}
  \bibinfo{person}{Goldie Nejat}.} \bibinfo{year}{2019}\natexlab{}.
\newblock \showarticletitle{How robots influence humans: A survey of nonverbal
  communication in social human--robot interaction}.
\newblock \bibinfo{journal}{\emph{International Journal of Social Robotics}}
  \bibinfo{volume}{11}, \bibinfo{number}{4} (\bibinfo{year}{2019}),
  \bibinfo{pages}{575--608}.
\newblock


\bibitem[Sz{\'e}kely et~al\mbox{.}(2019)]%
        {szekely2019spontaneous}
\bibfield{author}{\bibinfo{person}{{\'E}va Sz{\'e}kely},
  \bibinfo{person}{Gustav~Eje Henter}, \bibinfo{person}{Jonas Beskow}, {and}
  \bibinfo{person}{Joakim Gustafson}.} \bibinfo{year}{2019}\natexlab{}.
\newblock \showarticletitle{Spontaneous Conversational Speech Synthesis from
  Found Data.}. In \bibinfo{booktitle}{\emph{INTERSPEECH}}.
  \bibinfo{pages}{4435--4439}.
\newblock


\bibitem[{Tobii Pro AB}(2014)]%
        {TobiiProLab}
\bibfield{author}{\bibinfo{person}{{Tobii Pro AB}}.}
  \bibinfo{year}{2014}\natexlab{}.
\newblock \bibinfo{title}{Tobii Pro Lab}.
\newblock \bibinfo{howpublished}{Computer software}.
\newblock
\urldef\tempurl%
\url{http://www.tobiipro.com/}
\showURL{%
\tempurl}


\bibitem[Vinciarelli et~al\mbox{.}(2011)]%
        {Vinciarelli2012}
\bibfield{author}{\bibinfo{person}{Alessandro Vinciarelli},
  \bibinfo{person}{Maja Pantic}, \bibinfo{person}{Dirk Heylen},
  \bibinfo{person}{Catherine Pelachaud}, \bibinfo{person}{Isabella Poggi},
  \bibinfo{person}{Francesca D'Errico}, {and} \bibinfo{person}{Marc
  Schroeder}.} \bibinfo{year}{2011}\natexlab{}.
\newblock \showarticletitle{Bridging the gap between social animal and unsocial
  machine: A survey of social signal processing}.
\newblock \bibinfo{journal}{\emph{IEEE Transactions on Affective Computing}}
  \bibinfo{volume}{3}, \bibinfo{number}{1} (\bibinfo{year}{2011}),
  \bibinfo{pages}{69--87}.
\newblock


\bibitem[Wolfert et~al\mbox{.}(2022)]%
        {Wolfert_Robinson_Belpaeme_2021}
\bibfield{author}{\bibinfo{person}{Pieter Wolfert}, \bibinfo{person}{Nicole
  Robinson}, {and} \bibinfo{person}{Tony Belpaeme}.}
  \bibinfo{year}{2022}\natexlab{}.
\newblock \showarticletitle{A review of evaluation practices of gesture
  generation in embodied conversational agents}.
\newblock \bibinfo{journal}{\emph{IEEE Transactions on Human-Machine Systems}}
  (\bibinfo{year}{2022}).
\newblock


\bibitem[Yoon et~al\mbox{.}(2019)]%
        {Yoon_Ko_Jang_Lee_Kim_Lee_2019}
\bibfield{author}{\bibinfo{person}{Youngwoo Yoon}, \bibinfo{person}{Woo-Ri Ko},
  \bibinfo{person}{Minsu Jang}, \bibinfo{person}{Jaeyeon Lee},
  \bibinfo{person}{Jaehong Kim}, {and} \bibinfo{person}{Geehyuk Lee}.}
  \bibinfo{year}{2019}\natexlab{}.
\newblock \showarticletitle{Robots learn social skills: End-to-end learning of
  co-speech gesture generation for humanoid robots}. In
  \bibinfo{booktitle}{\emph{2019 International Conference on Robotics and
  Automation (ICRA)}}. IEEE, \bibinfo{pages}{4303--4309}.
\newblock


\end{thebibliography}
\appendix





\end{document}